\documentclass[acmsmall,screen]{acmart}

\AtBeginDocument{%
  \providecommand\BibTeX{{%
    \normalfont B\kern-0.5em{\scshape i\kern-0.25em b}\kern-0.8em\TeX}}}

\usepackage[utf8]{inputenc}
\usepackage[T1]{fontenc}
\usepackage{microtype}
\usepackage{algorithmic}
\usepackage{graphicx}
\usepackage{textcomp}
\usepackage{xcolor}
\usepackage{multirow, graphics}
\usepackage{listings}
\usepackage{diagbox}
\usepackage{subfig}
\usepackage{balance}
\usepackage{adjustbox}
\usepackage{colortbl}
\usepackage{framed}
\usepackage{booktabs}
\usepackage{indentfirst}
\usepackage{makecell,tcolorbox}
\usepackage[framemethod=TikZ]{mdframed}
\usepackage[linesnumbered,ruled,vlined]{algorithm2e}
\usepackage{tcolorbox}
\tcbuselibrary{breakable}
\usepackage{tikz}
\usepackage{pgfplots}
\usepackage{subcaption}
\pgfplotsset{compat=1.18}
\usepgfplotslibrary{groupplots}
\usepackage{pgfplots}
\usepgfplotslibrary{groupplots}
\pgfplotsset{compat=1.18}

\definecolor{sodagrey}{RGB}{220,220,220}
\definecolor{sodagreen}{RGB}{190,225,180}
\definecolor{sodayellow}{RGB}{245,205,80}
\definecolor{sodared}{RGB}{210,130,130}
\definecolor{sodablue}{RGB}{150,175,215}

\def\renderinsertedpart{}

\ifx\renderinsertedpart\undefined

\newenvironment{insertedpartenv}{}{}
\else

\fi

\begin{document}

\title{TraceDev: A Traceability-Driven Multi-agent  Framework for Requirement-to-Code Development}

\author{Mingyu Chen}
\orcid{0009-0003-7041-7194}
\authornotemark[2]
\email{220110123@stu.hit.edu.cn}
\affiliation{
  \institution{Harbin Institute of Technology, Shenzhen}
  \state{Guangdong}
  \country{China}
}
\author{Yakun Zhang}
\authornotemark[2]
\authornote{Corresponding author; $^\dagger$ Equal contribution.}
 \email{zhangyk@hit.edu.cn}
 \orcid{0009-0009-2377-3114}
\affiliation{
  \institution{Shenzhen Key Laboratory of Internet Information Collaboration, Harbin Institute of Technology, Shenzhen, China  and State Key Laboratory of Novel Software Technology, Nanjing University, Nanjing}
  \country{China}
}

\author{Zihao Xie}
\orcid{0009-0008-8682-4153}
\email{25s151153@stu.hit.edu.cn}
\affiliation{
  \institution{Shenzhen Key Laboratory of Internet Information Collaboration, Harbin Institute of Technology, Shenzhen}
  \state{Guangdong}
  \country{China}
}

\author{Yixing Luo}
\email{luoyi\_xing@126.com}
\orcid{0000-0001-6223-5217}
\affiliation{
  \institution{Beijing Institute of Control Engineering, Beijing}
  \country{China}
}

\author{Jinrui Xu}
\email{220111026@stu.hit.edu.cn}
\orcid{0009-0001-7295-1388}
\affiliation{
  \institution{Harbin Institute of Technology, Shenzhen}
  \state{Guangdong}
  \country{China}
}

\author{Cuiyun Gao}
\email{gaocuiyun@hit.edu.cn}
\orcid{0000-0003-4774-2434}
\affiliation{
  \institution{Harbin Institute of Technology, Shenzhen}
  \state{Guangdong}
  \country{China}
}

\author{Kaiqi Zhao}
\email{zhaokaiqi@hit.edu.cn}
\orcid{0000-0002-0984-1629}
\affiliation{
 \institution{Shenzhen Key Laboratory of Internet Information Collaboration, Harbin Institute of Technology, Shenzhen}
  \state{Guangdong}
  \country{China}
}

\author{Yunming Ye}
\email{yeyunming@hit.edu.cn}
\orcid{0000-0002-1807-8581}
\affiliation{
 \institution{Shenzhen Key Laboratory of Internet Information Collaboration, Harbin Institute of Technology, Shenzhen}
  \state{Guangdong}
  \country{China}
}




\begin{abstract}
In modern software development, the rapid advancement of Large Language Models (LLMs) has made the end-to-end transformation of Natural Language Requirements (NLRs) into executable repository-level code increasingly feasible. However, existing approaches typically rely on simplified instructions (e.g., single-sentence descriptions), failing to reflect complex software development scenarios. Moreover, they lack explicit requirement traceability mechanisms, making it difficult to precisely align and validate generated code against original requirements.
To address these limitations, we propose TraceDev, a multi-agent framework for automated software development grounded in use cases that contain multiple functional points and complex semantics. TraceDev employs five role-specific agents, including a Requirement Refiner, Designer, Developer, Tester, and Validator. Notably, the Validator Agent constructs and maintains a heterogeneous traceability graph that links requirements, design models, and code artifacts for interacting with the preceding four agents. The traceability graph maintains consistency across various artifacts and
serves as a structured context for efficient memory management, supporting reliable repository-level code generation. We evaluate TraceDev on two widely used datasets (including 125 use cases) compared with two state-of-the-art approaches. On the ETOUR dataset, TraceDev achieves a success rate of 53.63\%, outperforming baseline approaches by up to 186.63\%. A similar trend is observed on the SMOS dataset, where TraceDev attains a success rate of 56.82\%, exceeding baseline approaches by up to 340.80\%. These results demonstrate the effectiveness of TraceDev in repository-level code generation from requirements. 
\end{abstract}


\setcopyright{cc}
\setcctype{by}
\acmDOI{10.1145/3832171}
\acmYear{2026}
\acmJournal{PACMSE}
\acmVolume{3}
\acmNumber{ISSTA}
\acmArticle{ISSTA080}
\acmMonth{10}
\acmSubmissionID{issta26main-p753-p}
\received{2026-01-30}
\received[accepted]{2026-04-16}







\graphicspath{{figures/}}
\newcommand{\figref}[1]{Figure~\ref{#1}}
\newcommand{\tabref}[1]{Table~\ref{#1}}
\newcommand{\secref}[1]{Section~\ref{#1}}

\newcommand{\toolName}[1]{\textbf{TraceDev\raisebox{-0.7ex}{\Huge\textbf{#1}}}}

\newcommand{\toolNameplain}[1]{TraceDev\raisebox{-0.7ex}{\Huge #1}}

\newcommand{\toolNameSmall}{TraceDev}
\newcommand{\datasetetour}{ETOUR}
\newcommand{\datasetsmos}{SMOS}
\newcommand{\oururl}{\url{https://github.com/ISSE-Lab/ISSTA2026-TraceDev}}

\maketitle
\newcommand{\revise}[1]{\textcolor{blue}{[revise: #1]}}

\newcommand{\ie}{\mbox{\emph{i. e.,\ }}}

\section{Introduction}\label{sec:introduction}
With the increasing complexity of modern software systems, traditional software development faces significant challenges~\cite{complexityone,complexitytwo,complexitythree,chen2025towards}. The long chain from requirements analysis and system design to coding and testing is not only time-consuming but also error-prone~\cite{winters2020software,zadenoori2025large}. In recent years, the rapid development of Large Language Models (LLMs), represented by Gemini~\cite{team2023gemini} and DeepSeek~\cite{guo2024deepseek}, has brought a revolutionary paradigm shift to the field of software engineering~\cite{gao2025current,xi2025rise,huynh2025large,chen2025smaller,chen2025deep}. These techniques make the vision of ``Requirements as Code'' increasingly feasible~\cite{zadenoori2025large}. Consequently, the focus of both industry and academia is shifting from simple code completion to higher-level tasks (i.e., the end-to-end transformation of Natural Language Requirements (NLRs) into \emph{repository-level code generation})~\cite{wu2024autogen,li2022competition,chen2021evaluating,qian2024chatdev, hong2024metagpt}.

Existing studies mainly focus on building automated frameworks based on multi-agent collaboration~\cite{qian2024chatdev, hong2024metagpt}, in which different agents play different roles and collaborate to simulate the entire Software Development Life Cycle (SDLC).
While these studies have demonstrated progress in the end-to-end transformation of Natural Language Requirements (NLRs) into repository-level code, they still face  the following two limitations.

\textbf{\textit{1) Simplified Development Scenarios.}}
In real-world software development, requirements are typically documented in a structured and fine-grained form, among which use cases are one of the most prevalent formats for requirements specification. 
An empirical study surveying requirement specifications across 12 companies finds that 33\% of interviewees adopt use cases as requirement specifications, demonstrating the widespread application of use cases in real-world software projects~\cite{practice_2023}.
Such requirements described in use cases typically contain \emph{multiple functional points and rich semantic constraints}. However, existing studies mainly focus on development scenarios with simple semantics and structures, such as code generation based on single-sentence natural language instructions or incomplete code snippets (e.g., HumanEval~\cite{chen2021evaluating} and MBPP~\cite{mbpp}). \emph{These research scenarios differ substantially from real-world requirements in terms of task complexity.}
    
\textbf{\textit{2) Absence of Traceability.}} During the development and maintenance of software systems, there are numerous software artifacts (e.g., requirements, design models, and source code) that need to be managed~\cite{fuchss2025lissa}. Traceability allows requirements to be traced to related design models and code, facilitating a clear understanding of artifact relationships and improving software maintainability~\cite{zadenoori2025large}. Existing studies rarely consider the critical role of traceability throughout the Software Development Life Cycle. 
\emph{Without traceability, semantic deviations and functional omissions may arise during cross-stage transformations of various artifacts}, such as requirements, design models, and code implementations.
This undermines the completeness, reliability and maintainability of code generation in complex development scenarios.

\textbf{Our work. }To address the preceding two limitations, we propose TraceDev, a novel \textbf{Trace}ability-driven multi-agent Framework for requirements-to-code \textbf{Dev}elopment.
Specifically, TraceDev distinguishes itself from existing studies through two key features:
\textbf{\textit{1) Suitable for complex development scenarios.}}
TraceDev is specifically designed for complex development scenarios, leveraging use cases composed of multiple function points as requirements rather than simple descriptions. It achieves repository-level code generation\footnote{
  Repository-level code generation in this paper refers to generating a system-level implementation for a complete use-case, consisting of multiple collaborating code files.
} by comprehending the full scope of the use case, rather than generating isolated functions.
\textbf{\textit{2) Traceability Graph construction and utilization. }}TraceDev constructs a heterogeneous traceability graph linking requirements, design models, and code artifacts during the development process. The traceability graph enables automated detection of unimplemented or incorrectly implemented requirements, \emph{supporting reliable repository-level code generation in complex development scenarios}. Meanwhile, it also
serves as a structured context for managing software artifacts, \emph{providing an efficient memory management mechanism for Multi-Agent Systems}.

To support the preceding two features, TraceDev adopts a multi-agent framework composed of role-specific agents. It employs five agents:  Requirement Refiner, Designer, Developer, Tester, and Validator, which are responsible for requirement refinement, design modeling, code implementation, testing, and requirement traceability, respectively. These agents work together to simulate the Software Development Life Cycle (SDLC). 
In particular, the Validator Agent constructs and maintains a traceability graph to preserve consistency across various artifacts. It also provides an efficient shared memory for all agents to support continuous validation and iterative refinement.



To comprehensively assess the effectiveness of TraceDev,
we evaluate TraceDev using two widely adopted datasets in the Requirement Engineering (RE) domain, i.e., ETOUR and SMOS~\cite{coest}, which include 125 use cases. We compare TraceDev with two state-of-the-art approaches, ChatDev~\cite{qian2024chatdev} and MetaGPT~\cite{hong2024metagpt}. The evaluation covers three complementary dimensions: automatic evaluation, human evaluation, and code-level statistical analysis. For automatic evaluation, two metrics are used: \textit{semantic-coverage rate}, which measures the coverage of semantic implementation, and \textit{success rate}, which measures the functional correctness of implementation. 
The experimental results show that TraceDev is effective in the end-to-end transformation of Natural Language Requirements (NLRs) into repository-level code. On the ETOUR dataset, TraceDev attains a semantic-coverage rate of 71.72\%, surpassing ChatDev by 51.66\% and MetaGPT by 75.14\%, and a success rate of 53.63\%, exceeding ChatDev by 129.19\% and MetaGPT by 186.64\%. A similar trend is observed on the SMOS dataset, where TraceDev achieves a success rate of 56.82\%, surpassing the baselines by up to 340.80\%. 

The main contributions of our work are summarized as follows:
\begin{itemize}
    \item 
    To the best of our knowledge, this work is the first to introduce the traceability graph into code generation. 
    The traceability graph maintains consistency across various software artifacts and serves
as a structured context for efficient memory management.
    Furthermore, our work focuses on complex development scenarios, where requirements are described in use cases with multiple functional points and rich semantic constraints.
    
    \item 
    We propose TraceDev, a  traceability-driven multi-agent framework with five specialized roles: Requirement Refiner, Designer, Developer, Tester, and Validator. 
    These agents collaboratively automate the Software Development Life Cycle.
    Notably, the Validator Agent maintains a traceability graph and interacts with the other agents, supporting consistency validation and efficient memory management throughout the development process.
    
    \item 
    We evaluated TraceDev on two representative datasets. Experimental results demonstrate that TraceDev outperforms the state-of-the-art baseline approaches, effectively generating high-quality repository-level code based on requirements.
    
\end{itemize}

\section{Motivational Example}\label{sec:example}

In this section, we use \datasetetour{}, an electronic tourism system, as a motivating example to illustrate the characteristics of  requirements in complex development scenarios and the challenges arising from the absence of requirement traceability in software development.


\subsection{Inadequacy of Existing Benchmarks for Complex Development Scenarios}

\begin{figure*}[t]
	\center
 \includegraphics[scale = 0.48]{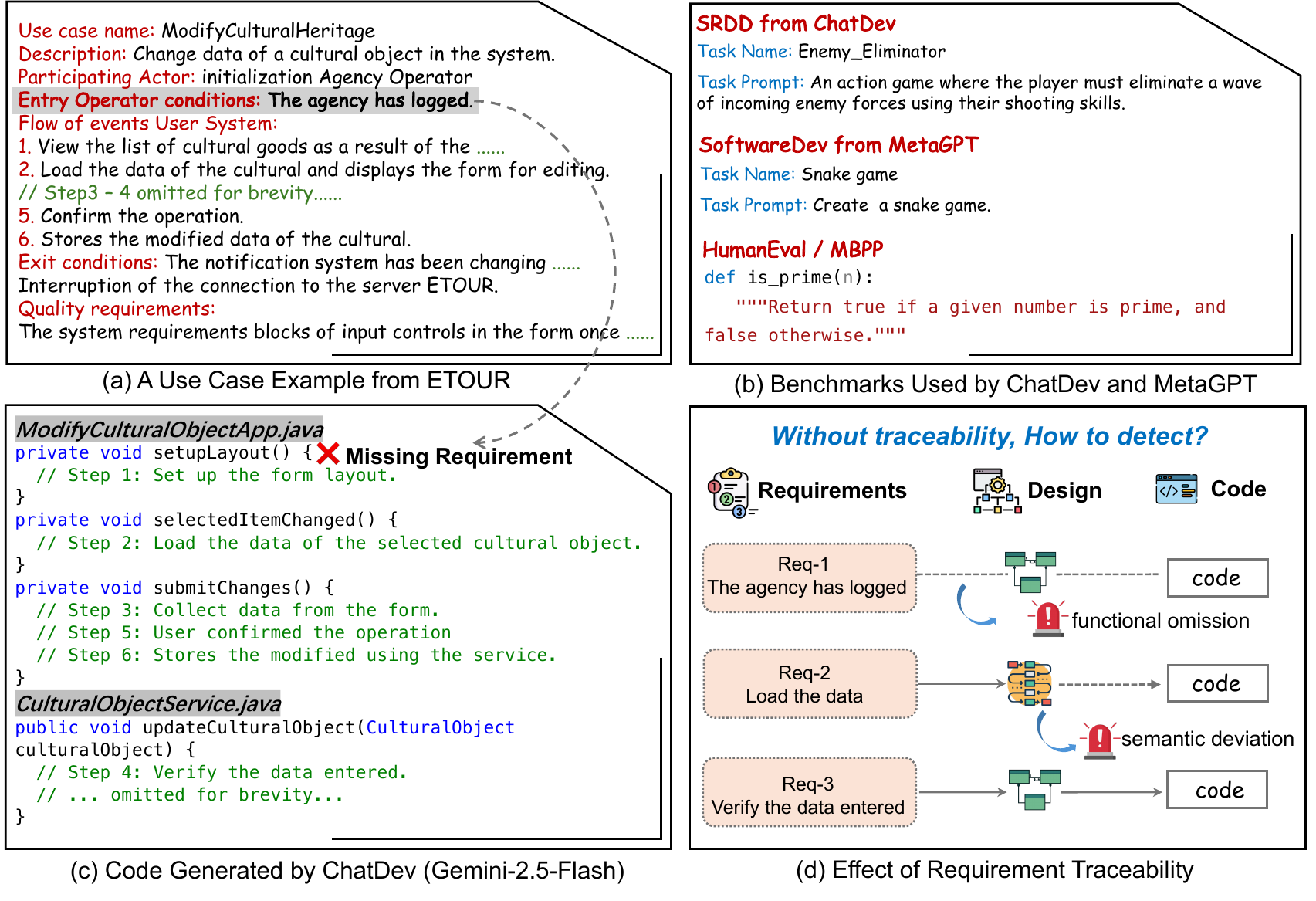}
 \vspace{-4mm}
	\caption{The motivation example.}
	\label{fig:motivation-example}
\vspace{-5mm}
\end{figure*}

In the \datasetetour{} system, requirements are documented in the form of use cases. \figref{fig:motivation-example}(a) presents a representative use case from the ETOUR system, namely \texttt{ModifyCulturalHeritage}. This use case is described in a structured and fine-grained form, consisting of \texttt{a participating actor, entry conditions, a detailed flow of events, exit conditions, and quality requirements}. A single use case contains multiple functional points and rich semantic constraints.
Such use cases are prevalent in complex development scenarios because they provide a structured format for precisely representing functions, flows, and constraints~\cite{12768}.

However, existing approaches mainly focus on simplified development  scenarios\cite{hong2024metagpt,qian2024chatdev}.
As illustrated in \figref{fig:motivation-example}(b), current state-of-the-art approaches are evaluated on benchmarks that typically feature simple and coarse-grained inputs. 
For instance, ChatDev~\cite{qian2024chatdev} introduces the SRDD (Software Requirement Description Dataset) benchmark, which consists of task-oriented software requirement descriptions. An example from the SRDD benchmark is \texttt{Enemy\_Eliminator}, described as: ``\texttt{An action game where the player must eliminate a wave of incoming enemy forces using their shooting skills}''. Similarly, MetaGPT~\cite{hong2024metagpt} is evaluated on the SoftwareDev benchmark,  which mainly contains simplified software development requirements (e.g., \texttt{Create a Snake Game}). 
In addition, public benchmarks such as HumanEval~\cite{chen2021evaluating} and MBPP~\cite{mbpp} mainly consist of function-level programming tasks, with each task covering only a single functionality.

Both simplified requirement descriptions, such as SRDD and SoftwareDev, as well as programming tasks like HumanEval and MBPP, focus only on single functionalities or coarse-grained tasks.
However, these simplified tasks are insufficient to reflect the nature of requirements in complex development scenarios.
\textbf{Consequently, there is an urgent need to develop approaches capable of handling complex development scenarios with real-world requirements (e.g., comprehensive and lengthy use cases).}


\subsection{Absence of Requirement Traceability}

\figref{fig:motivation-example}(c) presents code generated by ChatDev~\cite{qian2024chatdev} using Gemini-2.5-Flash~\cite{comanici2025gemini}, based on the use case example from the \datasetetour{} dataset shown in \figref{fig:motivation-example}(a). 
Although the generated code appears to follow the main execution flow, it fails to fully cover all specified requirements. For instance, the entry condition ``\texttt{The agency has logged in}'' is omitted in the implementation. 
This observation indicates that existing LLM-based agents fail to ensure requirement completeness.

The root cause of this issue lies in the absence of requirement traceability throughout the generation process. 
As shown in \figref{fig:motivation-example}(d), when requirement traceability is absent, the generated code cannot be reliably traced back to individual requirements, making it difficult to detect missing or incorrectly implemented requirements. 
Especially during cross-stage transformations among heterogeneous software artifacts such as requirements, design, and code, semantic deviations or functional omissions are prone to occur, thereby severely undermining the completeness and correctness of the implementation.
In contrast, when requirement traceability is maintained, each requirement can be systematically traced and verified against the generated code, enabling the identification of missing or incorrect implementations and supporting targeted refinement.

\textbf{Therefore, it is essential to establish requirement traceability to ensure completeness, correctness, and reliability in real-world software development.}

\section{TraceDev}\label{sec:ftdroid}

In this section, we propose \toolNameSmall{}, a traceability-driven multi-agent framework that enables end-to-end repository-level code generation from Natural Language Requirements (NLRs) described in use cases. We first present the overview of \toolNameSmall{} and then describe its details.


\begin{figure}[t]
	\center
 \includegraphics[width=14cm]{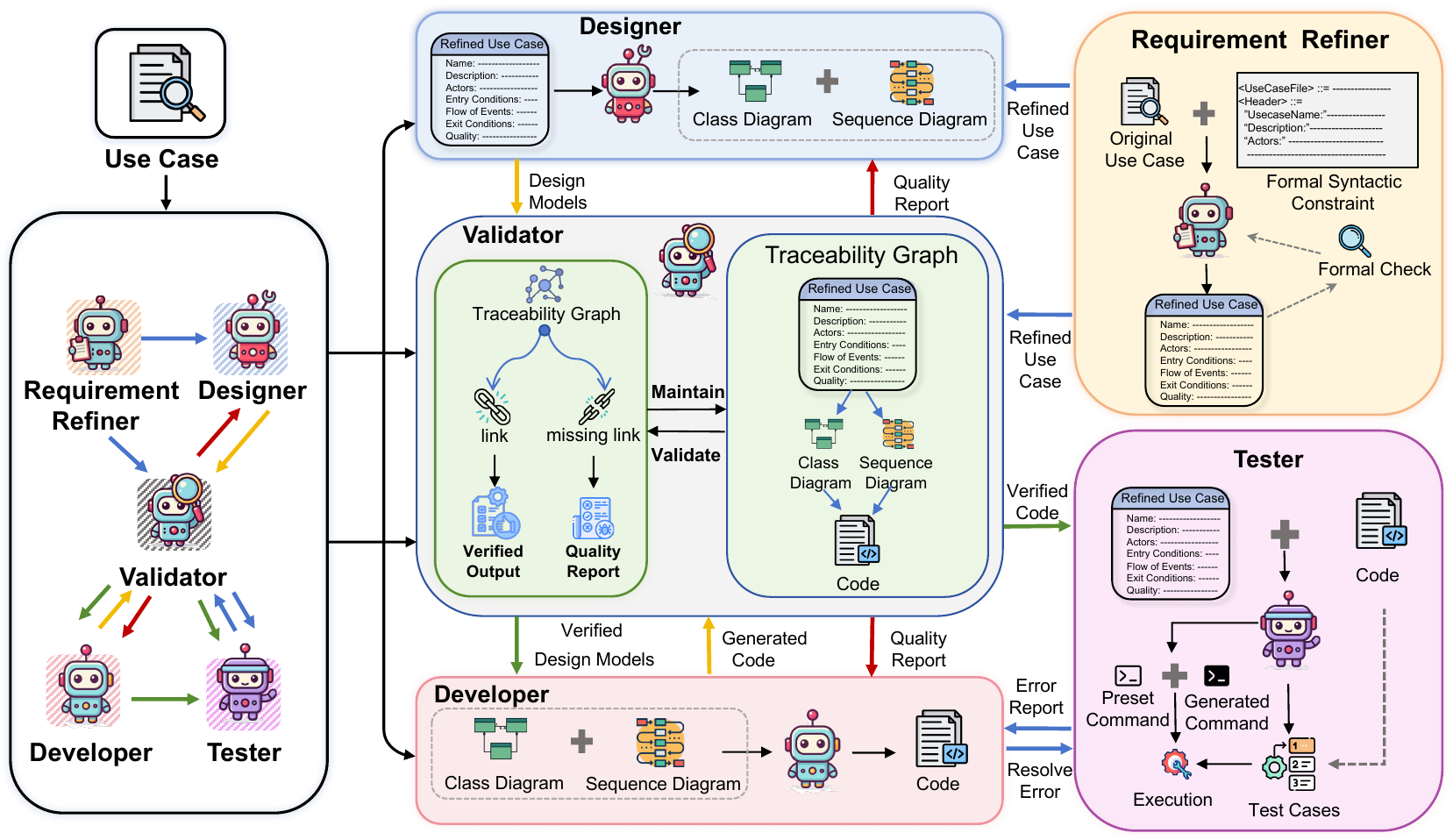}
	\caption{An overview of \toolNameSmall{}.}
	\label{fig:overview}
\end{figure}

\subsection{Overview}

As illustrated in \figref{fig:overview}, \toolNameSmall{} takes a use case as input, which describes the system’s behavior under various conditions in response to requests from a user~\cite{jacobson1993objectusecase}.
\toolNameSmall{} simulates the Software Development Life Cycle by leveraging multiple LLM-based agents.
These agents collaboratively transform the refined use case into a series of software artifacts through specialized roles, including a \textbf{Requirement Refiner} for eliminating terminology inconsistency and semantic ambiguity in use case, a \textbf{Designer} for generating class and sequence diagrams, a \textbf{Developer} for code generation, a \textbf{Tester} for test case construction and execution, and a \textbf{Validator} for maintaining the traceability graph to assess consistency and completeness across different artifacts.

Through the multi-agent collaboration, \toolNameSmall{} automates the end-to-end software development process, encompassing requirement refinement, system design, code generation, and testing.
It iteratively enhances the reliability of artifacts through test execution and the maintenance of a traceability graph,  ultimately supporting more automated and reliable software development

\begin{figure*}[t]
	\center
 \includegraphics[scale = 0.45]{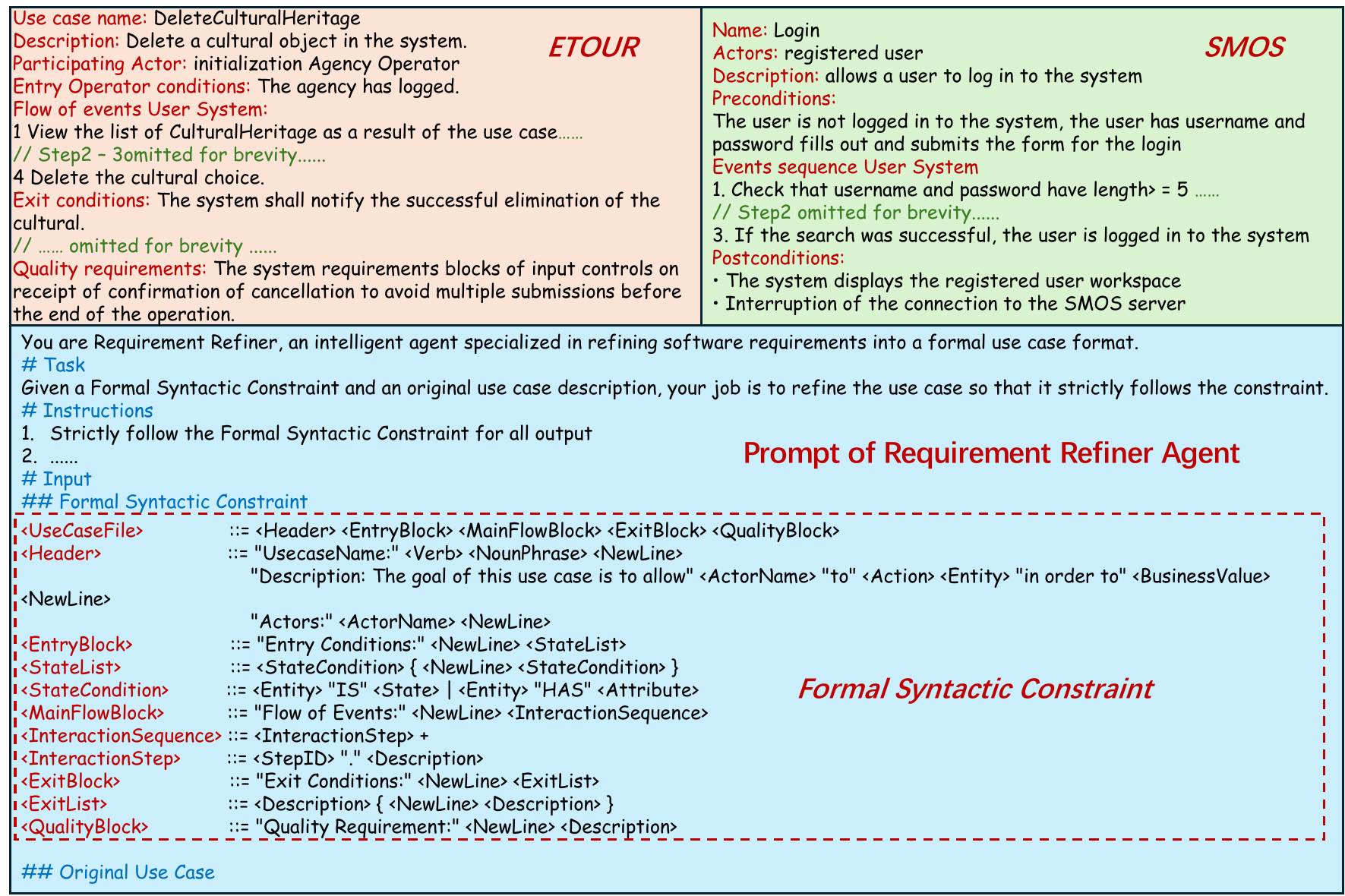}
	\caption{A comparison of ETour and SMOS use cases, along with the Requirement Refiner Agent’s prompt that contains a formal syntactic constraint.}
	\label{fig:refiner}
\end{figure*}

\subsection{Requirement Refiner Agent}

In \toolNameSmall{}, the Requirement Refiner Agent is responsible for resolving terminological inconsistencies and semantic ambiguities in original use cases. As illustrated in \figref{fig:overview}, it acts as a preprocessor that takes original use cases as input and produces refined use cases with unified terminology and explicit semantics.

We analyze ETOUR and SMOS, two widely recognized datasets in the requirements domain, and identify \textbf{\emph{the following challenges}}:
\begin{itemize}
    \item \textbf{Challenge 1: Terminology Inconsistency.} 
    Although the use case specification proposed by Alistair Cockburn has been widely recognized as a standard~\cite{cockburn2008writingusecase}, real-world software systems commonly exhibit variations in the naming of core use case terminology (e.g., names, actors, preconditions, and main flows).
    As illustrated in \figref{fig:refiner}, when describing the main execution flow, SMOS uses \texttt{Events Sequence}, whereas ETOUR uses \texttt{Flow of Events}. Similarly, actor fields are labeled as \texttt{Actors} by SMOS and \texttt{Participating Actor} by ETOUR, respectively. Such terminology inconsistencies limit the applicability of automated tools, particularly when requirements are extracted from a use case as nodes using terminology-based rules or regular expressions.
    \item \textbf{Challenge 2: Semantic Ambiguity.} Even if terminology is aligned, a more severe issue lies in the ambiguity of natural language descriptions. As illustrated in \figref{fig:refiner}, in the \texttt{Login} use case from SMOS , Step 1 states, “\texttt{Check that username and password have length $\ge$ 5...}”.  This sentence omits the acting subject, making it ambiguous whether the validation is performed by the system or by the user. Similarly, in the ETOUR, Step 1 “\texttt{View the list of CulturalHeritage...}” lacks an explicit subject (implicitly the actor). Without resolving these ambiguities, the downstream Designer Agent might incorrectly assign responsibilities (e.g., generating system operations within user classes), leading to logical errors in the generated class diagrams and sequence diagrams.
\end{itemize}

To address the two challenges, the Refiner Agent performs the following two key mechanisms.

\textbf{(1) Schema Alignment and Subject Inference.} 
As illustrated in \figref{fig:refiner}, a formal syntactic constraint based on Alistair Cockburn’s detailed use case specification~\cite{cockburn2008writingusecase} is injected into the LLM prompt. Guided by this constraint, the LLM performs two tasks in the refinement process. First, it interprets the semantics of the original use case and normalizes the terminology into a standardized schema. For example, terminologies such as \texttt{"Entry Operator Conditions"} or \texttt{"Preconditions"} are unified into \texttt{"Entry Conditions"}.
Second, the LLM infers missing subjects from contextual information and rewrites sentences into atomic $〈Subject + Verb + Object〉$ forms. For example, “Check that...” is transformed into “The system checks that...”. 
By performing schema normalization and subject inference during use case refinement, the Requirement Refiner Agent ensures that the refined use cases are both semantically consistent and grammatically complete.

\textbf{(2) Iterative Formal Verification.} The Refiner Agent employs a syntactic parser to automatically check whether the refined use case satisfies the predefined Formal Syntactic Constraint. If a violation is detected, a feedback loop is initiated to prompt the LLM to regenerate until the output passes the formal check. This mechanism ensures that only structurally valid and unambiguous use cases are passed to downstream agents.

By introducing the Requirement Refiner Agent, we effectively bridge terminological gaps between projects and eliminate semantic ambiguities, ensuring that the downstream Designer Agent can perform modeling based on precise and unambiguous requirements.

\subsection{Designer Agent}

As illustrated in \figref{fig:overview}, 
in \toolNameSmall{}, the Designer Agent takes the refined use case as input and generates design models, including class diagrams and sequence diagrams.

In real-world software development, software design constitutes a bridge between the requirements and the implementation~\cite{ullrich2025requirements}. Directly converting natural language requirements into programming code often results in code that is poorly structured and difficult to maintain. 
To address the challenge, we introduce a Designer Agent, which allows \toolNameSmall{} to establish the architectural skeleton and interaction logic before delving into implementation details. 
By decoupling structural definition from code implementation, the Designer Agent can specialize in capturing the structure and interactions of the system, improving the modularity and maintainability of the system.

\begin{figure}[t]
    \centering
    \begin{minipage}[b]{0.45\textwidth}  
        \centering
        \includegraphics[width=\textwidth]{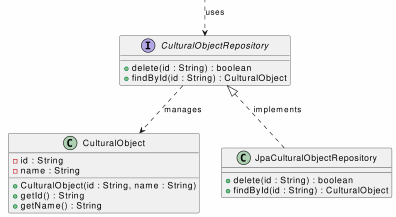}
        \caption{A class diagram generated by \toolNameSmall{} using Gemini-2.5-Flash. Due to space limitations, only a subset of the diagram is shown. The diagram is based on the use case from \figref{fig:refiner}.}
        \label{fig:left}
    \end{minipage}
    \hfill  
    \begin{minipage}[b]{0.48\textwidth}
        \centering
        \includegraphics[width=\textwidth]{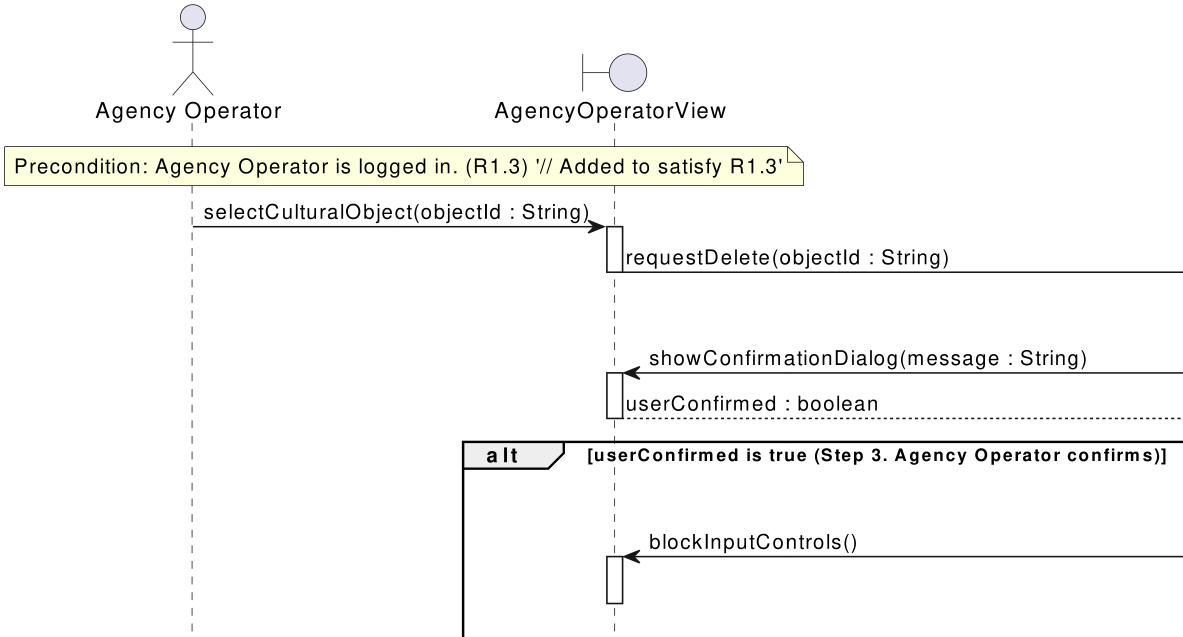}
        \caption{A sequence diagram generated by \toolNameSmall{} using Gemini-2.5-Flash. Due to space limitations, only a subset of the diagram is shown. The sequence diagram is based on the use case from \figref{fig:refiner}.}
        \label{fig:right}
    \end{minipage}
\end{figure}

To provide a complete description of the system’s structure and behavior, the Designer Agent generates two complementary UML diagrams in PlantUML code:

\begin{itemize}
    \item \textbf{Class Diagram: } As illustrated in \figref{fig:left}, a class diagram defines the static architecture of the system. The LLM analyzes noun phrases within the use case to identify \texttt{Classes}, \texttt{Attributes}, and \texttt{Methods}, as well as \texttt{Inheritance} and \texttt{Association Relationships}. This ensures that the generated code adheres to object-oriented structural principles.
    \item \textbf{Sequence Diagram:} As illustrated in \figref{fig:right}, a sequence diagram depicts the runtime interactions between objects. The LLM analyzes the flow of events in the use case and derives ordered interactions and message flows, thereby ensuring that the designed architectural behavior aligns with the intended execution order of the use case logic.
\end{itemize}

The Designer Agent bridges the abstraction gap between natural language requirements and code, enabling the LLM to establish a clear architectural skeleton and interaction logic in advance, thereby improving the modularity, structure, and maintainability of the generated code.


\newcommand\mycommfont[1]{\footnotesize\ttfamily\textcolor{teal}{#1}}
\SetCommentSty{mycommfont}

\begin{algorithm}[t]
\caption{Collaborative Code Generation and Verification Process}
\label{alg:dev_test_loop}
\SetKwInOut{Input}{Input}
\SetKwInOut{Output}{Output}
\newcommand{\DevGen}[1]{\mathrm{GenerateCode}(#1)}
\newcommand{\TestGen}[2]{\mathrm{GenerateTests}(#1,#2)}
\newcommand{\Execute}[2]{\mathrm{Execute}(#1,#2)}
\newcommand{\DevFix}[3]{\mathrm{SelfCorrect}(#1,#2,#3)}

\Input{Refined Use Case $\mathcal{U}$, Design Artifacts $\mathcal{D}$ (Class and Sequence Diagrams), Maximum retry limit $K$}
\Output{Verified Executable Code $\mathcal{C}_{final}$}
Initialize retry counter $iter \leftarrow 0$\;

$\mathcal{C} \leftarrow \DevGen{\mathcal{D}}$\;

$\mathcal{T} \leftarrow \TestGen{\mathcal{U}, \mathcal{C}}$\;

$iter \leftarrow 0$\;
\While{$iter < K$}{
    $pass, \mathcal{L}_{log} \leftarrow \Execute{\mathcal{C}, \mathcal{T}}$\;
    
    \If{$status = true$}{ \tcp*[h]{All tests passed}
        \Return $\mathcal{C}$ 
    }
    
    $\mathcal{F} \leftarrow$ Extract failure information from $\mathcal{L}_{log}$\;
    $\mathcal{C} \leftarrow \DevFix{\mathcal{C}, \mathcal{F}, \mathcal{D}}$ 
    $iter \leftarrow iter + 1$\;
}

\Return $\mathcal{C}$ \tcp*[h]{Return latest version with warning if max retries reached}
\end{algorithm}

\subsection{Developer \& Tester Agents}

As illustrated in ~\figref{fig:overview} and Algorithm ~\ref{alg:dev_test_loop}, after the Designer Agent produces design artifacts, the Developer Agent is responsible for code implementation, while the Tester Agent is responsible for test case generation and execution. Based on the execution results, the two agents collaboratively perform feedback-driven self-correction.

\textbf{Code Generation Based on Design Artifacts}.
The Developer Agent takes the class diagrams and sequence diagrams in PlantUML code produced by the Designer Agent as input. These design artifacts are then injected into the prompt, guiding the LLM to generate code whose static structure and dynamic behavior are consistent with the provided UML models (Line 2 in Algorithm ~\ref{alg:dev_test_loop}).

\textbf{Test Case Generation and Execution.}
Following code generation, the Tester Agent takes the refined use case and the generated code as input and injects them into a prompt for the LLM. Based on this input, the LLM generates unit test cases that cover both main execution paths and exception handling scenarios. Subsequently, using LLM-generated command lines combined with manually preset commands, the Tester Agent automatically compiles the generated code and executes the test cases (Line 2 \& 6 in Algorithm ~\ref{alg:dev_test_loop}).

\textbf{Self-Correction via Feedback Loop.}
Following test execution, if any failures are observed, the Tester Agent extracts execution feedback, including error messages and test outcomes (Line 9 in Algorithm ~\ref{alg:dev_test_loop}).
This feedback is forwarded to the Developer Agent, which regenerates corrected code (Line 10 in Algorithm ~\ref{alg:dev_test_loop}).
This feedback loop iterates with a predefined maximum number of retries to prevent unbounded correction (Line 5 in Algorithm ~\ref{alg:dev_test_loop}).


\subsection{Validator Agent}

In the software development life cycle simulated by TraceDev, the Refiner, Designer, and Developer Agents sequentially produce diverse artifacts, including refined use cases, design models, and code. However, semantic deviations or functional omissions can easily arise during cross-stage transformations of these heterogeneous artifacts.
To address this challenge, the Validator Agent is introduced to construct a traceability graph to maintain the completeness and consistency of the end-to-end development process.
As shown in \figref{fig:overview},  the Validator Agent constructs a heterogeneous \textbf{traceability graph} that links the refined use cases from the Requirement Refiner Agent, the design models from the Designer Agent, and the generated code from the Developer Agent.

The traceability graph enables automatic identification of missing links, where requirements lack relevant design elements or design elements lack concrete code implementations.
Meanwhile, it also serves as a structured context for managing software artifacts. \emph{Consequently, the traceability graph serves as a robust foundation for large-scale code generation and an efficient memory management mechanism for Multi-Agent Systems.}
Based on this graph, the Validator Agent identifies inconsistencies between artifacts, generates a quality report, and passes targeted feedback to the Designer or Developer Agents, thereby guiding iterative refinement.




\SetKwInput{KwInput}{\textbf{input}}
\SetKwInput{KwOutput}{\textbf{output}}

\begin{algorithm}[t]
\caption{Traceability Graph Construction}
\label{alg:traceability}

\KwInput{Refined Use Case $\mathcal{U}$, Design Models $\mathcal{D}$, Generated Code $\mathcal{C}$}
\KwOutput{Traceability Graph $\mathcal{G}(V, E)$}


$V_{req} \leftarrow \text{ExtractNodes}(\mathcal{U})$ \tcp*[h]{Actors, Entry/Exit Conditions, Steps in Flow of Events and Quality Requirement}

$V_{design} \leftarrow \text{ExtractNodes}(\mathcal{D})$ \tcp*[h]{Classes, Methods, Relations, Messages}

$V_{code} \leftarrow \text{ExtractNodes}(\mathcal{C})$ \tcp*[h]{Code Files}

Initialize $\mathcal{G}$ with $V \leftarrow V_{req} \cup V_{design} \cup V_{code}$ and $E \leftarrow \emptyset$\;

\ForEach{node $u \in V_{req}$}{
    \tcp{Query LLM to identify correlation via constructed prompts}
    $d_{target} \leftarrow \text{LLM\_SemanticMatch}(u, V_{design})$\;
    
    \If{$d_{target} \neq \emptyset$}{
        \tcp{create edge from requirement to design models}
        $E \leftarrow E \cup \{(u \to d_{target})\}$\;
    }
}

\ForEach{node $d \in V_{design}$}{
    \tcp{verify syntactic implementation via AST parsing}
    $c_{target} \leftarrow \text{AST\_SyntacticMatch}(d, V_{code})$\;
    
    \If{$c_{target} \neq \emptyset$}{
        \tcp{create edge from design models to generated code}
        $E \leftarrow E \cup \{(d \to c_{target})\}$\;
    }
}

\Return{$\mathcal{G}(V, E)$}\;
\end{algorithm}

\begin{figure}[t]
	\center
 \includegraphics[width=13cm]{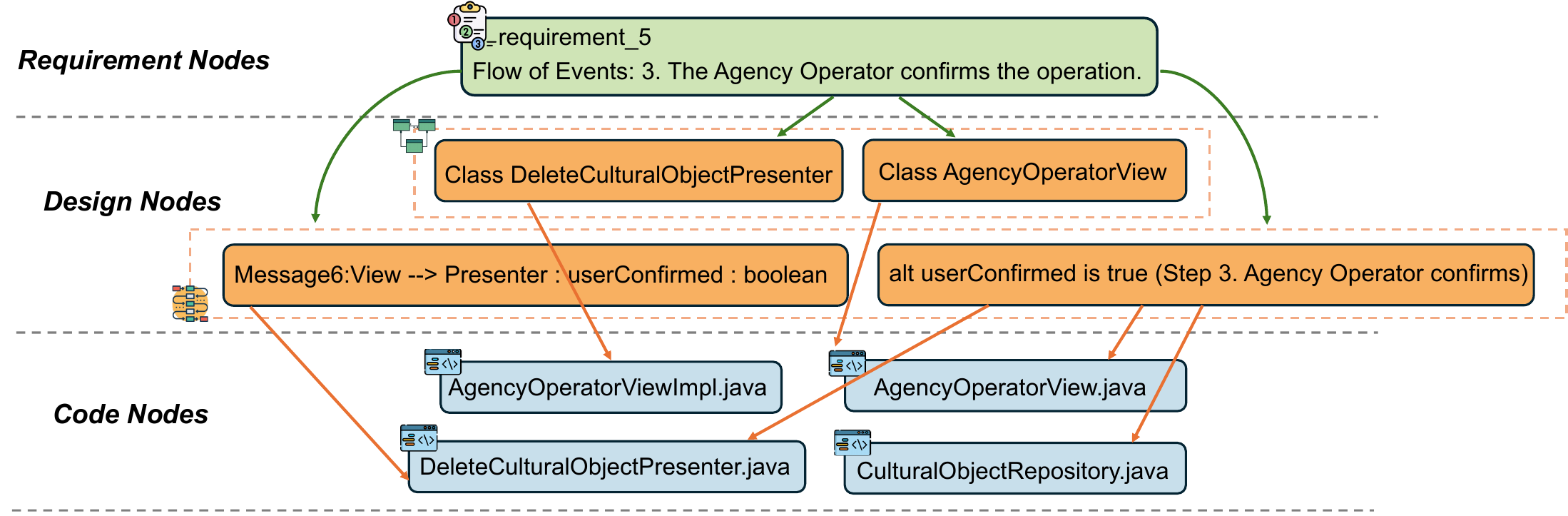}
	\caption{A traceability graph generated by TraceDev using Gemini-2.5-Flash. Due to space limitations, only a subset of the graph is shown for clarity. The traceability graph is based on the ETOUR use case from \figref{fig:refiner}.}
	\label{fig:graph}
\end{figure}

\subsubsection{Traceability Graph.} 
We first introduce how the traceability graph is constructed and then highlight its key benefits in repository-level code generation.

\textbf{Traceability Graph Construction.}
To assess traceability completeness, the Validator Agent constructs a graph over heterogeneous software artifacts. 
The construction process of the traceability graph is detailed in Algorithm~\ref{alg:traceability}. \figref{fig:graph} presents a traceability graph generated by TraceDev using Gemini-2.5-Flash based on the ETOUR use case from \figref{fig:refiner}.
Formally, the traceability graph is defined as a directed graph $G = (V, E)$, where the node set $V$ represents artifact entities, and the edge set $E$ represents traceability links.

Specifically, the node set $V$ is composed of three subsets extracted from different lifecycle stages. For requirements ($\mathcal{U}$), we employ regular expressions to parse the refined use cases to extract semantic entities such as \texttt{actors, entry/exit conditions, flow steps, and quality requirements }(Line 1 in Algorithm ~\ref{alg:traceability}). For design models ($\mathcal{D}$), we analyze the structural representations to identify classes, methods, relations, and messages (Line 2 in Algorithm ~\ref{alg:traceability}). Finally, for generated code ($C$),  each generated code file is modeled as a node (Line 3 in Algorithm ~\ref{alg:traceability}). The set $V$ is the union of these extracted subsets, as shown in line 4 of Algorithm ~\ref{alg:traceability}.

The establishment of traceability between artifacts is divided into two main phases:

\textit{(1) Requirement-to-Design Tracing.} As shown in lines~5--8 of Algorithm ~\ref{alg:traceability}, the LLM analyzes each requirement element in the refined use case and identifies relevant design entities from the Class and Sequence Diagrams. If a relevant logical representation is found in the design models, a directed edge is established from the requirement node to the design node.

\textit{(2) Design-to-Code Tracing. }As shown in lines~9--12 of Algorithm ~\ref{alg:traceability}, this phase verifies whether the design elements are implemented in the generated code. Specifically, the Validator Agent uses Abstract Syntax Tree (AST) parsing~\cite{tree-sitter} to analyze the generated code, identifying concrete implementations related to the entities in the design models. When a matching definition or method call is found, a directed edge is established from the design node to the code node.

\textbf{Memory Management Based on Traceability Graph.}
Beyond assessing traceability completeness, the traceability graph also serves as a structured context for managing software artifacts. In repository-level code generation, artifacts (e.g., requirements, design models, and code) are often large in scale, making it costly to directly incorporate entire artifacts into the model context.
To address this challenge, the traceability graph is used to provide a compressed representation of the context. Specifically, the Validator Agent extracts only the key semantic and structural elements from requirements, design models, and code, and organizes them in the form of nodes and edges. In this way, large-scale artifacts are transformed into a compact and interpretable graph-based representation. This representation preserves the core dependencies among artifacts while significantly reducing the overall context size.

\subsubsection{Iterative Refinement via Multi-Agent Interaction.} 
To ensure traceability completeness, the Validator Agent interacts with the Designer Agent and the Developer Agent during the design and coding phases, respectively. Specifically, the Validator Agent automatically traverses the traceability graph to identify missing edges, including both requirement-to-design and design-to-code links. These missing links are then injected into prompts for the LLM, which generates a quality report highlighting potential defects. The report is subsequently provided as targeted feedback to the Designer Agent and the Developer Agent to guide corrections in the design and code artifacts. This process is iterated with a maximum number of iterations enforced to prevent infinite loops, enabling progressive improvement of traceability completeness across various artifacts.

\section{Experimental Setup}\label{sec:evaluation_setup}
To comprehensively evaluate TraceDev, we answer research questions focusing on three key perspectives.
First, we assess the effectiveness of TraceDev by comparing it with two state-of-the-art multi-agent approaches on two representative datasets, using Gemini-2.5-Flash~\cite{comanici2025gemini} and DeepSeek-V3.2~\cite{liu2025deepseek} as foundation models.
These two models represent state-of-the-art coding and reasoning performance during the experimental period while offering complementary cost-efficiency.
Second, we analyze the contribution of each agent within TraceDev through ablation studies to understand their individual impacts on overall performance. 
Third, we evaluate the efficiency of TraceDev in comparison with the two baseline approaches. Details are as follows.

\textbf{RQ1:} How effective is TraceDev compared with state-of-the-art multi-agent approaches?


\textbf{RQ2:} How does each agent contribute to the overall performance of TraceDev? 

\textbf{RQ3:} How efficient is TraceDev compared to baseline approaches?

\begin{table}[t]
    \centering
    \caption{The datasets used for evaluation.}
    \label{tab:datasets}
    \begin{tabular}{lcccccc}
        \toprule
        \multirow{2}{*}{Dataset} & \multirow{2}{*}{Domain} & \multicolumn{2}{c}{Requirement} 
        & {Evaluation} \\
        \cmidrule(lr){3-4} 
        \cmidrule(lr){5-5}
         & & Type & Number 
         & Test Case\\
        \midrule
        eTour & Tourism & Use Case & 58 
        & 962\\
        SMOS & Education & Use Case & 67 
        & 1179\\
        \bottomrule
    \end{tabular}
\end{table}
\vspace{-2mm}
\begin{figure*}[t]
	\center
 \includegraphics[scale = 0.4]{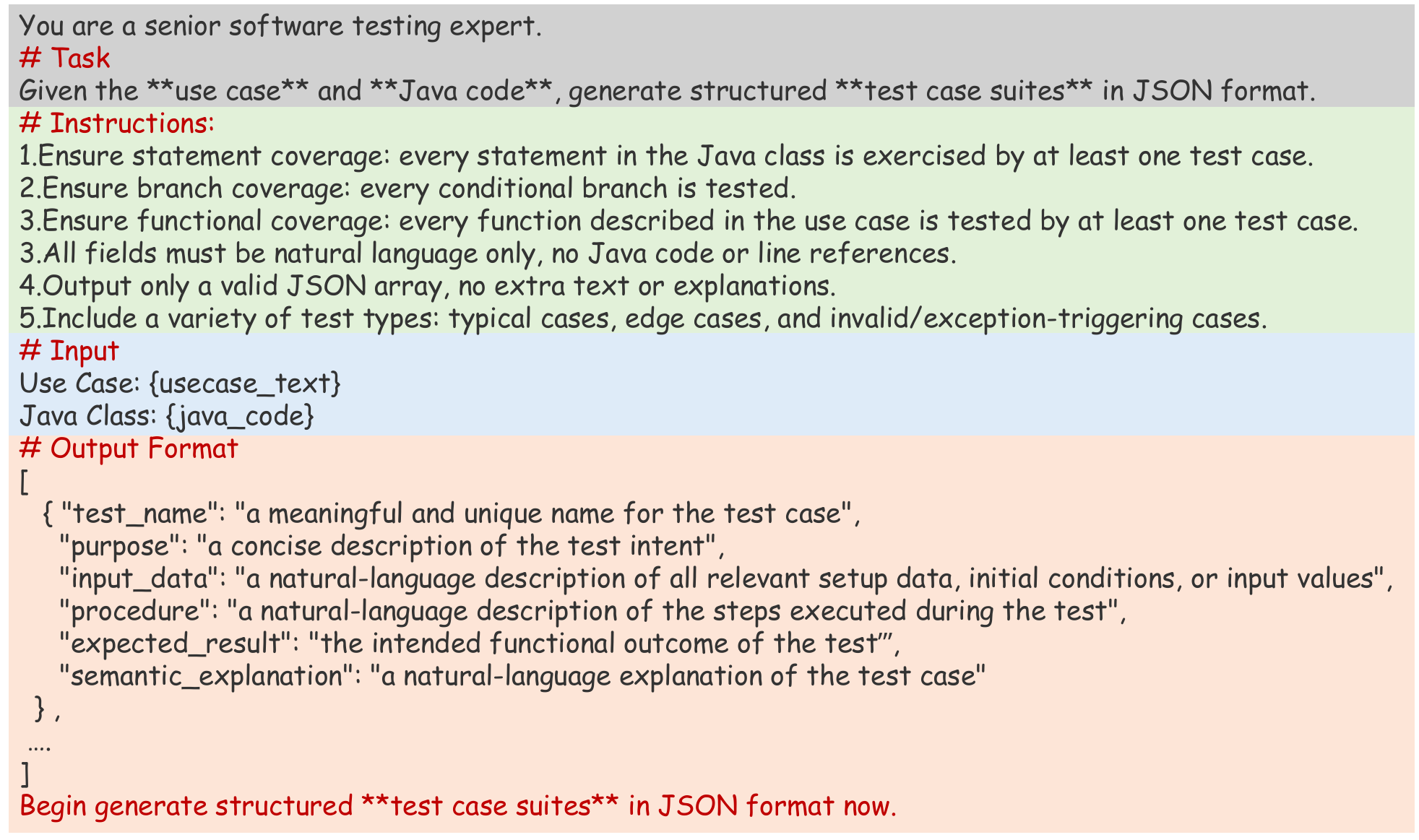}
	\caption{The prompt for LLM-guided test case generation using use cases and ground-truth code.}
	\label{fig:test_case_prompt}
    \vspace{-3mm}
\end{figure*}

\subsection{Datasets}\label{sec:datasets}

Our evaluation is conducted using two real-world system datasets, eTour and SMOS. Table \ref{tab:datasets} represents a summary of the two datasets. The two selected datasets are provided by the Center of Excellence for Software \& Systems Traceability (CoEST)~\cite{coest} and cover different application domains, including education and tourism. 
Each dataset contains multiple use cases, and each use case is implemented by multiple code files that together form a code repository.

Following prior works~\cite{luo2025rpg,zhang2025empowering}, we leverage the use cases and ground-truth implementations provided by the datasets.
Based on these, we construct evaluation tasks using test case generation to assess the semantic and functional consistency between the generated code and the ground-truth code.
Specifically, in the ETOUR and SMOS datasets, we first inject the use cases along with the relevant ground-truth code into the prompt (see \figref{fig:test_case_prompt}), guiding the LLM in generating test cases.
The generated test cases are designed to achieve statement and branch coverage and to test all functional points in the use case, providing a comprehensive verification of the system’s functional behavior.
To reduce the risk of test incorrectness, we retain the test cases that pass on ground-truth code.
In total, 962 test cases are generated for ETOUR and 1,179 for SMOS.

\subsection{Baselines}\label{sec:baselines}

To demonstrate the effectiveness of TraceDev, we compare it with the following baselines.


\begin{itemize}
    \item \textbf{ChatDev}~\cite{qian2024chatdev} is a communicative multi-agent framework that leverages a ``chat chain'' and role specialization to enable LLM agents to cooperatively execute core software lifecycle phases, including design, implementation, and testing.
    \item \textbf{MetaGPT}~\cite{hong2024metagpt} is a multi-agent framework that assigns agents professional roles (e.g., Product Manager, Architect, Engineer, and QA Engineer) and embeds standardized operating procedures (SOPs) into agent collaboration to guide software development workflows.
\end{itemize}

\subsection{Metrics}

To comprehensively evaluate the performance of TraceDev and the selected baselines, we focus on three complementary dimensions, which are automatic evaluation, human evaluation and statistical analysis. Details are as follows. 

\subsubsection{Automatic evaluation.} Following prior work~\cite{luo2025rpg}, we adopt two metrics to evaluate generated code from both \textbf{semantic presence} and \textbf{functional correctness}. 
The evaluation is conducted on the generated test cases, consisting of 962 tasks in eTour and 1,179 tasks in SMOS.
Details of the test case generation process can be found in Section ~\ref{sec:datasets}.
The evaluation metrics are as follows:

\begin{itemize}
    \item \textbf{Semantic-coverage Rate} measures how well functions are semantically implemented. This metric verifies whether the generated code logically implements the target functionality. 
    Specifically, we adopt a LLM-as-a-Judge mechanism. To mitigate the variability of single-round judgments, we perform three independent rounds of judgment by the LLM and determine the final decision through majority voting. If the majority of judgments indicate correctness, the generated code is considered to implement the target semantics; otherwise, it is considered incorrect. This approach effectively reduces occasional errors and hallucinations in model evaluation, ensuring the stability of semantic judgments.
    \item  \textbf{Success Rate} measures the functional correctness of  implementation. For test cases that pass semantic verification, we adapt and execute them against the generated code to evaluate functional correctness. This aligns the evaluation with realistic software testing practices.
\end{itemize}


Additionally, to ensure fairness and reproducibility of the evaluation, we develop automated execution scripts to drive the entire evaluation process. DeepSeek-V3.2~\cite{liu2025deepseek} is used as the foundation model for automated evaluation due to its high cost-effectiveness and computational efficiency.

\subsubsection{Human evaluation.}
While automated evaluation provides quantitative metrics, it remains insufficient to fully reflect the code’s executability and alignment with the requirements.
Therefore,
following prior works~\cite{hong2024metagpt,zhang2025empowering}, we use human evaluation to further assess the performance of the generated code, focusing on the following metric:

 The metric \textbf{Executability} evaluates whether the generated code is completely incorrect or meets the requirements to some extent. Each code sample is rated on a scale from 1 to 4:
    \begin{itemize}
        \item 1: The generated code fails to compile or cannot execute.
        \item 2: The generated code executes but may not meet most workflow requirements.
        \item 3: The generated code executes and mostly aligns with the requirements.
        \item 4: The code executes and perfectly satisfies all requirements.
    \end{itemize}

For human evaluation, we invite three experienced software developers with at least four years of programming experience to participate in the assessment. 
Each participant independently reviews the generated code and evaluates it according to a predefined set of criteria. To reduce individual bias, we provide detailed evaluation guidelines and aggregate the final results by averaging the scores across all evaluators.

\subsubsection{Statistical analysis.}
We analyze the scale and statistics of the generated code to distinguish minimal solutions from more complete, realistic implementations. The metrics are as follows:
\begin{itemize}
    \item \textbf{File Count:} the number of valid code files, reflecting the modularity of the code.
    \item \textbf{Normalized LOC:} effective lines of code after removing comments, docstrings, and blank lines, used to measure implementation size.
    \item \textbf{Code Token Count:} the number of tokens in the normalized code, measured using a standard tokenizer, reflecting the lexical complexity and richness of implementation details.
\end{itemize}

\subsection{Implementation Details}
\label{subsec:4}

For the baseline approaches, we directly run their publicly available open-source implementations, making only the necessary adaptations to the LLM API interfaces while keeping all other implementations and parameter settings unchanged to ensure a fair comparison.
For temperature configuration, we follow the original settings of each framework whenever available. Specifically, ChatDev uses a temperature of 0.2, TraceDev uses a temperature of 0 to improve reproducibility, and MetaGPT adopts its dynamic temperature mechanism. Following prior work~\cite{Checkeval}, we additionally set the temperature to 0 for DeepSeek-V3.2 during automated evaluation to reduce output randomness.

The maximum number of iterations for repairing missing traceability links is set to three, and up to five remediation attempts are allowed for test failures.
These values are determined empirically: we conduct experiments on a subset of the datasets (about 20\% of the use cases) and find that these settings achieve a good balance between repair effectiveness and computational cost.
All experiments are conducted on a server equipped with two Intel Xeon Silver 4310 CPUs, 1 TB of RAM, running Ubuntu 22.04.5 LTS, and six NVIDIA RTX A6000 GPUs each with 48 GB memory.

\section{evaluation}\label{sec:evaluation}

\subsection{Effectiveness of TraceDev(RQ1)}
\textbf{Experimental Design.} To evaluate the effectiveness of TraceDev, we compare TraceDev with the baselines (i.e., ChatDev~\cite{qian2024chatdev} and MetaGPT~\cite{hong2024metagpt}) on the ETOUR and SMOS datasets using Gemini-2.5-Flash~\cite{comanici2025gemini} and Deepseek-V3.2~\cite{liu2025deepseek}. For human evaluation, we randomly sample 20 use cases (10 cases from each dataset) to conduct a manual assessment of the generated code.

\textbf{Experimental Results.} Table ~\ref{table:rq1} shows that TraceDev outperforms the baselines  across various datasets and models. 
For human evaluation, we adopt Krippendorff’s alpha reliability coefficient~\cite{krippendorff2011computing} to measure the inter-rater agreement among multiple evaluators. The overall coefficient is 0.940, indicating excellent agreement.
We have three observations from Table ~\ref{table:rq1}.

\begin{table}[t]
\centering
\caption{Evaluation results of baselines and TraceDev on two datasets using Gemini-2.5-Flash and Deepseek-V3.2. 
``Cov.'' denotes semantic-coverage rate and ``Success'' denotes success rate.
``Files'', ``LOC'', and ``Tokens'' denote the number of code files, normalized lines of code, and code token count, averaged per use case.}
\label{table:rq1}
\resizebox{\linewidth}{!}{
\begin{tabular}{c|c|c|cccccc}
\toprule
\multirow{2}{*}{\textbf{Dataset}} 
& \multirow{2}{*}{\textbf{Model}} 
& \multirow{2}{*}{\textbf{Approach}} 
& \multicolumn{2}{c}{Automatic Evaluation} 
& {Human Evaluation} 
& \multicolumn{3}{c}{Statistical Analysis} \\
\cmidrule(lr){4-5}  \cmidrule(lr){6-6} \cmidrule(lr){7-9}
& & & Cov.(\%) & Success(\%) & Executability & Files & LOC & Tokens \\
\midrule
\multirow{3}{*}{ETOUR} 
& \multirow{3}{*}{Gemini-2.5-Flash} 
& ChatDev 
& 47.29\% & 23.40\% & 2.8  & 4.81 & 308.21 & 2274.47 \\
& & Metagpt 
& 40.95\% & 18.71\% & 2.26 & 4.88 & 253.00 & 1873.41 \\
& & \cellcolor{gray!20}\textbf{TraceDev} 
& \cellcolor{gray!20}\textbf{71.72\%} 
& \cellcolor{gray!20}\textbf{53.63\%} 
& \cellcolor{gray!20}\textbf{3.60} 
& \cellcolor{gray!20}12.35 & \cellcolor{gray!20}408.77 & \cellcolor{gray!20}3278.81 \\
\midrule
\multirow{3}{*}{SMOS} 
& \multirow{3}{*}{Gemini-2.5-Flash} 
& ChatDev 
& 42.49\% & 24.74\% & 2.46  & 5.57 & 342.63 & 2543.30 \\
& & Metagpt 
& 37.40\% & 12.89\% & 2.26  & 4.79 & 288.96 & 2171.52 \\
& & \cellcolor{gray!20}\textbf{TraceDev} 
& \cellcolor{gray!20}\textbf{70.05\%} 
& \cellcolor{gray!20}\textbf{56.82\%} 
& \cellcolor{gray!20}\textbf{3.66} 
& \cellcolor{gray!20}13.75 & \cellcolor{gray!20}427.37 & \cellcolor{gray!20}3350.27 \\
\midrule
\multirow{3}{*}{ETOUR} 
& \multirow{3}{*}{DeepSeek-V3.2} 
& ChatDev 
& 41.58\% & 16.52\% & 2.9  & 4.07 & 380.83 & 2720.16 \\
& & Metagpt 
& 37.01\% & 20.01\% & 2.46  & 3.60 & 463.64 & 3503.79 \\
& & \cellcolor{gray!20}\textbf{TraceDev} 
& \cellcolor{gray!20}\textbf{54.78\%} 
& \cellcolor{gray!20}\textbf{47.19\%} 
& \cellcolor{gray!20}\textbf{3.63} 
& \cellcolor{gray!20}15.11 
& \cellcolor{gray!20}416.02 
& \cellcolor{gray!20}2564.47 \\
\midrule
\multirow{3}{*}{SMOS} 
& \multirow{3}{*}{DeepSeek-V3.2} 
& ChatDev 
& 31.46\% & 16.80\% & 2.6  & 3.75 & 330.75 & 2377.90 \\
& & Metagpt 
& 33.41\% & 16.01\% & 2.03  & 2.53 & 450.94 & 3374.47 \\
& & \cellcolor{gray!20}\textbf{TraceDev} 
& \cellcolor{gray!20}\textbf{58.60\%} 
& \cellcolor{gray!20}\textbf{44.78\%} 
& \cellcolor{gray!20}\textbf{3.66} 
& \cellcolor{gray!20}14.76 
& \cellcolor{gray!20}403.45 
& \cellcolor{gray!20}2444.03 \\
\bottomrule\end{tabular}
}
\end{table}

\textbf{(1) In automatic evaluation, TraceDev demonstrates consistent advantages in semantic consistency and functional  correctness. }
Using Gemini-2.5-Flash, it achieves a semantic-coverage rate over 70\% and a success rate above 53\% on both datasets,
significantly outperforming the baselines. This superiority remains when using DeepSeek-V3.2. On the ETOUR dataset, TraceDev attains a success rate of 47.19\%, which is more than double the performance of ChatDev (16.52\%) and MetaGPT (20.01\%). A similar trend is observed on the SMOS dataset, where TraceDev maintains a success rate of 44.78\%, whereas the baselines struggle to exceed 17\%. These results indicate that TraceDev consistently achieves higher semantic consistency and functional correctness.


\textbf{(2) In human evaluation, TraceDev demonstrates superior and stable performance in executability.}
It achieves high  executability scores above 3.6 across all datasets and LLMs, where scores above 3 indicate successful execution and strong alignment with requirements. In contrast, the baselines generally score below 3.0. For instance, with Gemini-2.5-Flash on the SMOS dataset, TraceDev attains an executability score of 3.66, representing improvements of 48.8\% and 61.9\% over ChatDev and MetaGPT, respectively. When using DeepSeek-V3.2, TraceDev consistently maintains an executability score of 3.66, significantly outperforming MetaGPT, which achieves only 2.03.
\textbf{(3) In statistical analysis, TraceDev generates code that is generally larger in scale and higher in complexity.} 
Specifically, TraceDev produces more code files  compared to the baselines. Taking the ETOUR dataset as an example, TraceDev generates over 12 code files per use case, which is more than three times the number produced by the baseline approaches.
Regarding code volume, TraceDev generally produces high normalized Lines of code and code token counts. For example, with Gemini-2.5-Flash on the SMOS dataset, TraceDev achieves a high normalized LOC of 427.37 and a lexical complexity of 3,350.27 code tokens, substantially surpassing ChatDev and MetaGPT. These results indicate that TraceDev not only improves functional correctness but also tends to generate more complete and modular implementations, rather than simplified or minimal solutions. 
Notably, when using DeepSeek-V3.2, MetaGPT produces the largest code volume but only achieves a 16–20\% success rate, suggesting that the code quality is low despite its large volume.


\begin{tcolorbox}[breakable,width=\linewidth-2pt,boxrule=0pt,top=2pt, bottom=2pt, left=4pt,right=4pt, colback=gray!15,colframe=gray!15]
\textbf{Answer to RQ1:} TraceDev consistently outperforms baseline approaches in automatic evaluation, human evaluation, and code-level statistical analysis, generating more functionally correct, executable, and comprehensive code.
\end{tcolorbox}

\newcommand{\cbox}[2]{%
    \setlength{\fboxsep}{1.5pt}
    \colorbox{#1}{``#2''}%
}

\definecolor{sodatext}{HTML}{6A1B9A} 
\definecolor{sodagray}{HTML}{E8E8E8} 
\definecolor{sodapurple}{HTML}{B4A7D6}
\definecolor{sodagreen}{HTML}{AEE0AD} 
\definecolor{sodayellow}{HTML}{F1C844} 
\definecolor{sodared}{HTML}{EF9A9A}   
\definecolor{sodablue}{HTML}{90CAF9}  
\definecolor{sodatext-dark}{HTML}{471172}   
\definecolor{sodagray-dark}{HTML}{B0B0B0}   
\definecolor{sodapurple-dark}{HTML}{7D66A4} 
\definecolor{sodagreen-dark}{HTML}{70A970}  
\definecolor{sodayellow-dark}{HTML}{B78F2B}  
\definecolor{sodared-dark}{HTML}{A25555}    
\definecolor{sodablue-dark}{HTML}{5B81C6}   

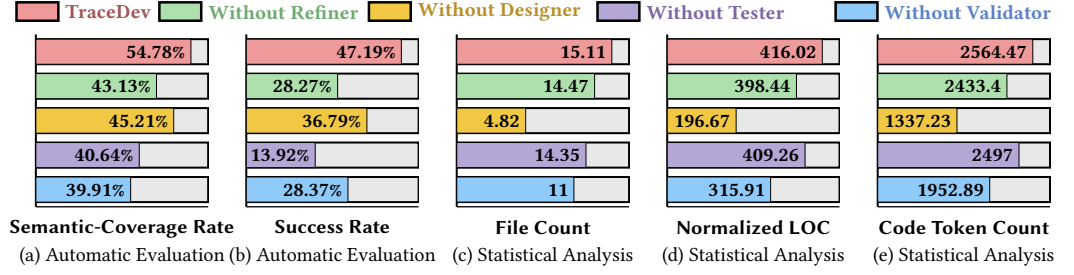
\begin{figure}[t]
\centering
\begin{tikzpicture}[scale=0.63, transform shape]
    \tikzstyle{legendbox} = [draw=black, line width=1.5pt, minimum width=2.50em, minimum height=1.1em]
    \tikzstyle{legendtext} = [text=sodatext, anchor=west, font=\Large\bfseries]
    
    \node (legend) at (10.5,4.1) { 
        \begin{tikzpicture}
            \node[legendbox, fill=sodared] (l1) at (0,0) {};
            \node[legendtext] at (l1.east) {\textcolor{sodared-dark}{TraceDev}};
            \node[legendbox, fill=sodagreen] (l2) at (3,0) {};
            \node[legendtext] at (l2.east) {\textcolor{sodagreen-dark}{Without Refiner}};
            \node[legendbox, fill=sodayellow] (l3) at (7.4,0) {};
            \node[legendtext] at (l3.east) {\textcolor{sodayellow-dark}{Without Designer}};
            \node[legendbox, fill=sodapurple] (l4) at (12.2,0) {};
            \node[legendtext] at (l4.east) {\textcolor{sodapurple-dark}{Without Tester}};
            \node[legendbox, fill=sodablue] (l5) at (17.2,0) {};
            \node[legendtext] at (l5.east) {\textcolor{sodablue-dark}{Without Validator}};
        \end{tikzpicture}
    };
    
    \begin{groupplot}[
        group style={
            group size=5 by 2,
            horizontal sep=0.8cm,
            vertical sep=1.7cm,
        },
        width=5.2cm, height=5.2cm,
        xbar,                  
        /pgf/bar shift=0pt,    
        /pgf/bar width=15pt,   
        xmin=0, xmax=100,
        ymin=0.5, ymax=5.5,    
        ytick={1,1.9,2.8,3.7,4.6},       
        yticklabels={},        
        xtick=\empty,          
        axis x line*=bottom,   
        axis y line*=left,      
        x axis line style={black, thick},
        y axis line style={black, very thick},
        ytick=\empty, 
        title style={at={(0.5,-0.1)}, anchor=north, font=\Large\sffamily},
        xlabel style={at={(0.5,-0.2)}, anchor=north, font=\Large},
        ]
        
        \newcommand{\addfourbars}[9]{
            \addplot[fill=sodagray, draw=black, very thick, forget plot] coordinates {
               (100, 5) (100, 4) (100, 3) (100, 2) (100, 1)
            };
            
            \addplot[fill=sodared, draw=black] coordinates {(90, 5)};
            \node[
                anchor=east,
                font=\Large\bfseries,
    text=black
            ] at (axis cs:90,5) {#5};
            
            \addplot[fill=sodagreen, draw=black] coordinates {(#1, 4)};
            \node[
                anchor=east,
                font=\Large\bfseries,
    text=black
            ] at (axis cs:#1,4) {#6};
            
            \addplot[fill=sodayellow, draw=black] coordinates {(#2, 3)};
            \node[
                anchor=east,
                font=\Large\bfseries,
    text=black
            ] at (axis cs:#2,3) {#7};
            
            \addplot[fill=sodapurple, draw=black] coordinates {(#3, 2)};
            \node[
                anchor=east,
                font=\Large\bfseries,
    text=black
            ] at (axis cs:#3,2) {#8};
            
            \addplot[fill=sodablue, draw=black] coordinates {(#4, 1)};
            \node[
                anchor=east,
                font=\Large\bfseries,
    text=black
            ] at (axis cs:#4,1) {#9};
        }


        \nextgroupplot[title={\textbf{Semantic-Coverage Rate}}, xlabel={(a) Automatic Evaluation}]
        \addfourbars{70}{80}{60}{55}{54.78\%}{43.13\%}{45.21\%}{40.64\%}{39.91\%}
        
        \nextgroupplot[title={\textbf{Success Rate}}, xlabel={(b) Automatic Evaluation}]
        \addfourbars{53}{70}{40}{59}{47.19\%}{28.27\%}{36.79\%}{13.92\%}{28.37\%}

        
        
        
        \nextgroupplot[title={\textbf{File Count}}, xlabel={(c) 
       Statistical Analysis}]
        \addfourbars{80}{40}{75}{68}{15.11}{14.47}{4.82}{14.35}{11}
        
        \nextgroupplot[title={\textbf{Normalized LOC}}, xlabel={(d) 
        Statistical Analysis}]
        \addfourbars{75}{40}{80}{60}{416.02}{398.44}{196.67}{409.26}{315.91}
        
        \nextgroupplot[title={\textbf{Code Token Count}}, xlabel={(e) 
        Statistical Analysis}]
        \addfourbars{75}{46}{82}{65}{2564.47}{2433.4}{1337.23}{2497}{1952.89}

    \end{groupplot}
\end{tikzpicture}
\caption{Ablation study. The \cbox{sodared}{red bar} denotes the original TraceDev, while the \cbox{sodagreen}{green bar}, \cbox{sodayellow}{yellow bar}, \cbox{sodapurple}{purple bar}, and \cbox{sodablue}{blue bar} denote TraceDev without the Refiner Agent, without the Designer Agent, without the Tester Agent, and without the Validator Agent, respectively.}
\label{fig:rq3}
\end{figure}


\subsection{Impact of Different Agents in TraceDev (RQ2)}
\textbf{Experimental Design.} We perform ablation studies to investigate the necessity of each agent in TraceDev, using the ETOUR dataset and the DeepSeek-V3.2 model.

\textbf{Experimental Results.} Figure ~\ref{fig:rq3} presents the ablation results under different agent-removal scenarios. Overall, we find that all agents are essential for achieving the best performance. 

\textbf{(1) The Validator Agent maintains requirement traceability and semantic completeness.}
Removing the Validator Agent leads to a noticeable drop in semantic-coverage rate, decreasing from 54.78\% to 39.91\%, the lowest among all variants. This indicates that without the Validator Agent identifying missing links through the traceability graph and facilitating iterative refinement, the generated code is prone to overlooking critical functionality.

\textbf{(2) The Tester Agent's iterative feedback maintains execution correctness.}
Removing the Tester Agent results in a dramatic drop in success rate to just 13.9\%, the lowest across all variants.
Notably, the lines of code ( LOC) generated by TraceDev without the Tester Agent are 409.26, very close to the 416.02 lines produced by the full framework. This indicates that simply generating more code is insufficient to guarantee execution correctness. Without the generate-test-refine loop, where the Developer and Tester agents iteratively collaborate to generate, test, and refine code, the code may appear complete but actually contains faults that prevent it from executing correctly.

\textbf{(3) The Designer Agent maintains structural modularity.}
In code statistics analysis, TraceDev without the Designer Agent shows the most significant reduction in code scale. The number of code files drops from 15.11 to 4.82, and the normalized line of code (LOC) decreases from 416.02 to 196.67, a reduction of more than half. This demonstrates the crucial role of the Designer Agent in architectural planning. Without its high-level guidance, the system tends to generate simple, non-modular single-file scripts, limiting its ability to handle complex tasks.

\begin{tcolorbox}[breakable,width=\linewidth-2pt,boxrule=0pt,top=2pt, bottom=2pt, left=4pt,right=4pt, colback=gray!15,colframe=gray!15]
\textbf{Answer to RQ2:} All agents are essential for TraceDev. Among all the agents, the Validator Agent maintains semantic completeness, the Tester Agent is critical for execution correctness, and the Designer Agent establishes structural modularity.
\end{tcolorbox}

\begin{table}[t]
\centering
\caption{Efficiency evaluation of TraceDev and baseline approaches.``Token Usage'' denotes the average number of tokens consumed per use case, ``Tim'' represents the average execution time. ``Files'', ``LoC'' and ``Code Token'' denote the number of code files, normalized lines of code, and code token count, respectively. ``Time effect.'' denotes the time effectiveness. The results for each approach are averaged across all use cases.}
\resizebox{\linewidth}{!}{
\begin{tabular}{c|c|c|cccccc}
\toprule
{\textbf{Dataset}} & {\textbf{Model}} & {\textbf{Approach}} & Token Usage & Time(s) & Files & Loc & Code Token  & Time effect.\\
\midrule
\multirow{3}{*}{ETOUR} & \multirow{3}{*}{Gemini-2.5-Flash} & ChatDev & 98,766 & 219.50 & 4.81 & 308.21 & 2274.47 & 0.11\\
 & & Metagpt & 219,758 & 169.30 & 4.88 & 253.00 & 1873.41 & 0.11\\
& & \cellcolor{gray!20}\textbf{TraceDev}
 & \cellcolor{gray!20}\textbf{109,541} 
 & \cellcolor{gray!20}\textbf{385.06} 
 & \cellcolor{gray!20}\textbf{12.35} 
 & \cellcolor{gray!20}\textbf{408.77} 
 & \cellcolor{gray!20}\textbf{3278.81}
 & \cellcolor{gray!20}\textbf{0.14} \\
\midrule
\multirow{3}{*}{SMOS} & \multirow{3}{*}{Gemini-2.5-Flash} & ChatDev & 109,601 & 261.94 & 5.57 & 342.63 & 2543.30 & 0.09\\
 & & Metagpt & 264,364 & 174.35 & 4.79 & 288.96 & 2171.52 & 0.07\\
& & \cellcolor{gray!20}\textbf{TraceDev}
 & \cellcolor{gray!20}\textbf{91,403} 
 & \cellcolor{gray!20}\textbf{325.56} 
 & \cellcolor{gray!20}\textbf{13.75} 
 & \cellcolor{gray!20}\textbf{427.37} 
 & \cellcolor{gray!20}\textbf{3350.27} 
 & \cellcolor{gray!20}\textbf{0.17} 
\\
\midrule
\multirow{3}{*}{ETOUR} & \multirow{3}{*}{DeepSeek-V3.2} & ChatDev & 71,456 & 587.84 & 4.07 & 380.83 & 2720.16 & 0.03\\
 & & Metagpt & 276,610 & 502.65 & 3.60 & 463.64 & 3503.79 & 0.04\\
& & \cellcolor{gray!20}\textbf{TraceDev}
 & \cellcolor{gray!20}\textbf{47,234} 
 & \cellcolor{gray!20}\textbf{535.26} 
 & \cellcolor{gray!20}\textbf{15.11} 
 & \cellcolor{gray!20}\textbf{416.02} 
 & \cellcolor{gray!20}\textbf{2564.47} 
  & \cellcolor{gray!20}\textbf{0.09} 
\\
\midrule
\multirow{3}{*}{SMOS} & \multirow{3}{*}{DeepSeek-V3.2} & ChatDev & 62,393 & 555.53 & 3.75 & 330.75 & 2377.90 & 0.03\\
 & & Metagpt & 193,318 & 327.33 & 2.53 & 450.94 & 3374.47 & 0.05\\
& & \cellcolor{gray!20}\textbf{TraceDev}
 & \cellcolor{gray!20}\textbf{47,503} 
 & \cellcolor{gray!20}\textbf{538.72} 
 & \cellcolor{gray!20}\textbf{14.76} 
 & \cellcolor{gray!20}\textbf{403.45} 
 & \cellcolor{gray!20}\textbf{2444.03} 
 & \cellcolor{gray!20}\textbf{0.08} 
\\
\bottomrule
\end{tabular}
}
\label{tab:efficiency}
\end{table}
\subsection{Efficiency of TraceDev(RQ3)}
\textbf{Experimental Design.} To assess efficiency, we evaluate execution time and token usage. Following prior works~\cite{zheng2025ars, lin2025plan}, we further calculate \emph{time effectiveness}, defined as the ratio of success rate to execution time, where a high value indicates higher resource utilization efficiency.

\textbf{Experimental Results.} As shown in Table ~\ref{tab:efficiency}, TraceDev maintains consistent efficiency across various datasets and models. Based on Table ~\ref{tab:efficiency}, we achieve the following observations.

\textbf{(1) In token consumption, TraceDev consistently demonstrates  excellent efficiency. } 
with DeepSeek-V3.2 on the ETOUR dataset, TraceDev achieves the lowest token usage across the two datasets, consuming only 47,234 tokens, which reduces token consumption by approximately 33\% compared to ChatDev and 83\% compared to MetaGPT.
Using Gemini-2.5-Flash, TraceDev consumes less than half the tokens of MetaGPT on the SMOS dataset. 
Notably, with Gemini-2.5-Flash on the ETOUR dataset, TraceDev’s total token consumption (109,541) is slightly higher than that of ChatDev (98,766). 
However, TraceDev demonstrates higher code productivity, requiring only 268 tokens per line of code on average, compared to approximately 320 tokens per line for ChatDev.


\textbf{(2) In execution time, TraceDev maintains  comparable efficiency.} 
Across all datasets and models, TraceDev achieves the highest time effectiveness.
Although TraceDev requires more execution time than the baselines when using Gemini-2.5-Flash, this is mainly because it generates a larger number of valuable code tokens.
For example, on the SMOS dataset with Gemini-2.5-Flash, TraceDev attains a time effectiveness of 0.17, outperforming ChatDev and MetaGPT by 88.89\% and 111.11\%. This indicates that the additional time is efficiently transformed into higher-quality code.


\textbf{(3) The time cost of TraceDev is primarily influenced by the Validator Agent,} which constructs a traceability graph across different artifacts and collaborates with the Designer Agent and Developer Agent through multiple iterations of “generate–validate–refine.” 
While introducing computational and interaction costs, this closed-loop mechanism ensures continuous self-correction, allowing TraceDev to enhance the reliability of the entire development process.



\begin{tcolorbox}[breakable,width=\linewidth-2pt,boxrule=0pt,top=2pt, bottom=2pt, left=4pt,right=4pt, colback=gray!15,colframe=gray!15]
\textbf{Answer to RQ3:} TraceDev is superior in token efficiency and maintains comparable execution time to baseline approaches, demonstrating overall strong efficiency.
\end{tcolorbox}

\subsection{Threats to Validity}
\textbf{External Validity.} A primary threat to external validity is the generalizability  to other datasets and foundation models. To mitigate the threat regarding the dataset, we utilize the ETOUR and SMOS datasets, which are widely recognized in the requirements engineering domain and simulate real-world industrial settings. 
Regarding model selection, we employed DeepSeek-V3.2~\cite{liu2025deepseek} and Gemini-2.5-Flash~\cite{comanici2025gemini} as the foundation LLMs. 
These two models are representative in terms of code generation and reasoning capabilities.

\textbf{Internal Validity.} A potential threat to internal validity arises from possible errors in our implementation, prompt design, and experimental execution. 
To mitigate these threats, we have open-sourced our entire implementation, experimental data, and all prompt templates, facilitating external validation and community feedback for continuous improvement. As for the human evaluation, we invited three experienced developers and provided them with a clear and standardized evaluation process to minimize subjective bias.

\textbf{Construct Validity. } A possible threat to construct validity is related to evaluation metrics. To mitigate this threat, we evaluate the effectiveness of the generated code from three complementary aspects rather than relying on a single metric. 
Specifically, these aspects include automatic evaluation, human evaluation, and statistical analysis.
All the evaluation metrics follow prior works~\cite{hong2024metagpt, zhang2025empowering,luo2025rpg}, providing a reliable and objective basis for assessing the quality of code produced by both TraceDev and the baseline approaches.

\section{discussion}
\textbf{Practical Implications.}
In requirement-driven software development, traceability plays a crucial role by linking requirements to their corresponding design and implementation~\cite{jin2025usertrace,borg2014recovering}.
It helps reduce the complexity of various development tasks, such as change impact analysis~\cite{8423658,guo2025natural}, bug localization~\cite{bugloc}, and software maintenance~\cite{6405269,maintain8}.
As multi-agent systems gradually gain the ability to generate large-scale software projects, human engineers responsible for subsequent maintenance often face the ``cold start'' problem~\cite{engideve}. The cold start refers to the difficulty of understanding and operating a new system due to a lack of sufficient context or prior knowledge. Without traceability that links requirements to implementations, understanding the logic of code or locating bugs becomes a high-cognitive-load task~\cite{292398,tracegood}. 
Existing code generation approaches mainly focus on functional correctness but neglect constructing traceability during generation. As a result, although the generated code may be useful, it lacks explicit links to the underlying requirements, which makes later maintenance difficult.
However, according to related study\cite{8020}, coding accounts for only about 20\% of the software lifecycle while maintenance occupies roughly 80\%. TraceDev provides clear guidance and support for this 80\% maintenance phase, substantially reducing the effort required for understanding and iterative development.

\begin{table}[t]
\centering
\caption{Evaluation results of TraceDev and baseline approaches using GPT-5-mini.}
\label{table:rq5}

\begin{tabular}{c|c|cc}
\toprule
\textbf{Model} 
& \textbf{Approach} 
& \multicolumn{2}{c}{Automatic Evaluation} \\
\cmidrule(lr){3-4}  
& & Semantic-coverage Rate(\%) & Success Rate(\%) \\
\midrule
\multirow{3}{*}{GPT-5-mini} 
& ChatDev &  43.06\% &  7.14\% \\
& Metagpt & 37.96\% &  25.51\% \\
& \cellcolor{gray!20}\textbf{TraceDev} 
& \cellcolor{gray!20}\textbf{64.69\%} 
& \cellcolor{gray!20}\textbf{51.84\%} \\
\bottomrule
\end{tabular}
\end{table}

\textbf{Generalizability Across LLMs.}
Our main experiments are conducted on Gemini-2.5-Flash and DeepSeek-V3.2, both of which demonstrate strong capabilities in code generation and reasoning during the evaluation period. To further assess the generalizability of our method, we additionally introduce an OpenAI model, \textit{GPT-5-mini}, for supplementary experiments. Given limited computational resources, we randomly sample 20\% subsets from the ETOUR and SMOS datasets for evaluation.
As shown in Table~\ref{table:rq1}, TraceDev consistently outperforms all baseline methods using GPT-5-mini, exhibiting stable and consistent performance improvements. Specifically, TraceDev achieves 64.69\% on the semantic-coverage rate and 51.84\% on the success rate, outperforming the best baseline by 21.63\% and 26.33\%, respectively.
These results indicate that the effectiveness of TraceDev does not depend on any specific large language model backbone, and it generalizes well across different model architectures with strong robustness.

\textbf{Impact of Traceability Graph Errors.}
Since TraceDev's iterative refinement relies on the traceability graph constructed by the Validator Agent, its accuracy directly affects the quality of the generated code. 
Traceability itself is a highly labor-intensive and costly task~\cite{rodriguez2023prompts,ruiz2023don}. Establishing perfectly accurate cross-artifact links in real-world development is inherently difficult, which is why prior work has extensively focused on automated traceability recovery~\cite{Retrospective2025Antoniol,grundy2002inconsistency,wan2025systematic,fuchss2025lissa}.
Accordingly, our goal is not a perfectly accurate graph, but an approximate structure sufficient to support iterative refinement. To this end, TraceDev employs LLM-based semantic matching for requirement-to-design links and AST-based syntactic matching for design-to-code links, with the Validator Agent automatically detecting missing links and providing feedback for repair. 
We acknowledge that incorrect links may trigger unnecessary modifications. Through error propagation analysis, we categorize such errors into two types, namely \textit{redundant links} and \textit{incorrect links}. \textit{Redundant links} only trigger additional checks without disrupting overall mappings, whereas \textit{incorrect links} may cause unnecessary modifications.
To mitigate such errors, the Validator Agent adopts a bounded repair mechanism, allowing up to three rounds of traceability repair and five rounds of test remediation (Section~\ref{subsec:4}), keeping the impact of error propagation within an acceptable range.

\textbf{Requirement Input Format.} 
TraceDev currently takes use cases as requirement inputs. The structured fields in use cases are used to drive the formal verification of the Requirement Refiner Agent and the extraction of requirement nodes in the traceability graph.
Although use cases are widely adopted in industrial practice~\cite{practice_2023}, requirements in real-world software development take diverse forms, including user stories, natural language requirement documents, and software requirements specifications (SRS), which differ in granularity and structure. For such non-use-case requirements, the preprocessing and traceability extraction mechanisms of TraceDev cannot be directly applied, and require redesigning the schema and parsing rules for each specific format.

\textbf{Lack of Traceability between Code and Tests. }The current traceability graph only covers three types of artifacts, namely requirement → design → code, and does not explicitly incorporate the test cases generated by the Tester Agent. As a result, the correspondence between test cases, requirements, and code is not yet explicitly modeled at the graph level. Nevertheless, TraceDev still leverages the feedback loop between the Tester Agent and the Developer Agent to iteratively validate and repair the generated code at the execution level, thereby improving its functional correctness to a certain extent.

\section{Related Work}\label{sec:related}
\label{sec:related}

\textbf{Requirement Granularity in Software Development. } In intelligent software development, the reliable transformation of requirements into executable code is an important research topic~\cite{chen2021evaluating, hou2024largerelatedwork,khan2025largerelatedworkone}.
Existing studies mainly focus on simple development scenarios, where code is generated from simple, coarse-grained natural language descriptions, such as single-sentence instructions~\cite{hong2024metagpt,qian2024chatdev} or short function descriptions~\cite{chen2021evaluating,mbpp,hou2024largerelatedwork}.
However, in real-world software development, requirements often exhibit high complexity and strict logical constraints~\cite{norheim2024challengesrelatedworkone,hou2024largerelatedwork}.
To manage this complexity, \textit{Use Cases} have been widely adopted as a standard specification form to structure requirements~\cite{jacobson1993objectusecase,cockburn2008writingusecase,Mencl2004,usecaseindur}.
Despite their prevalence in development practice, the transformation of structured use cases into executable code remains largely underexplored.

Previous studies usually rely on coarse-grained inputs that do not reflect real-world complex development scenarios.
In contrast, TraceDev adopts structured use cases as input to capture complex development scenarios, thereby more closely reflecting real-world software development.
\textbf{Traceability.}
Traceability plays a crucial role in software engineering by establishing and maintaining associations among various development artifacts, such as requirements, design models, test cases, and source code~\cite{jin2025usertrace,borg2014recovering}.
However, maintaining traceability manually is widely considered difficult in development practice due to its high costs in terms of time, budget, and effort~\cite{ruiz2023don}.
To reduce manual effort, researchers mainly focus on developing automated traceability recovery techniques, among which approaches based on information retrieval (IR) are the most prevalent~\cite{grundy2002inconsistency, Retrospective2025Antoniol, wan2025systematic}.
These approaches typically identify potential traceability links by measuring similarity between software artifacts such as source code and documentation, with representative techniques including the Vector Space Model (VSM) and probabilistic models~\cite{Retrospective2025Antoniol,wan2025systematic}.
With the advent of large language models (LLMs), recent studies have begun exploring their application in traceability recovery. These studies typically leverage techniques such as model fine-tuning~\cite{lin2021traceability}, prompt engineering~\cite{rodriguez2023prompts}, or retrieval-augmented generation (RAG)~\cite{fuchss2025lissa}.
Existing studies primarily focus on recovering traceability links after software artifacts (e.g., requirements, design models, and source code) are produced.
In contrast, TraceDev automatically constructs a Traceability Graph throughout the development process to support error detection, iterative refinement and long-term maintainability.

\textbf{Multi-Agent Collaboration in Software Development.} 
With the increasing application of LLMs in software engineering tasks, researchers have gradually recognized the limitations of single-model approaches in complex software development scenarios~\cite{zhang2023survey,e2025rag}. Real-world software development typically involves multiple stages, including requirements analysis, system design, code implementation, and test verification. A single LLM struggles to simultaneously satisfy the objectives and constraints of different stages~\cite{sommerville2011software,pressman2005software}. Consequently, recent studies have introduced multi-agent collaborative frameworks that simulate software development workflows through role specialization~\cite{hong2024metagpt,qian2024chatdev,li-etal-2025-codetree}. Following this paradigm, TraceDev adopts a multi-agent architecture, in which five specialized agents collaborate to simulate the entire software development lifecycle.

However, existing multi-agent frameworks lack mechanisms to address semantic deviations or functional omissions that can easily arise during cross-stage transformations in the software development lifecycle.
To address this challenge, TraceDev constructs a traceability graph, explicitly maintaining traceability between software artifacts (e.g., requirements, design models and code) produced throughout the development process. The graph helps maintain consistency between software artifacts and enhances software reliability.


\section{Conclusion}\label{sec:conclusion}
\label{sec:conclusion}

In this paper, we propose TraceDev, a traceability-driven multi-agent framework, which aims to achieve high-quality end-to-end transformation from use cases to executable repository-level code.  TraceDev employs five agents, including Requirement Refiner, Designer, Developer, Tester, and Validator, which are respectively responsible for requirement refinement, system design, code generation, testing, and  traceability validation. Throughout the development process, the Validator Agent constructs and maintains a heterogeneous Traceability Graph that links requirements, design models, and code artifacts, thereby supporting more reliable software development. We evaluate TraceDev on two real-world requirement datasets (including
125 use cases) and compare it with two state-of-the-art approaches, ChatDev and MetaGPT.
Experimental results demonstrate the superior performance of TraceDev. In future work, we plan to extend TraceDev to a wider range of LLMs, additional
programming languages, and optimize our prompt template.

\section{Data Availability}\label{sec:dataavail}
We release our source code and data at \oururl{}.


\bibliographystyle{ACM-Reference-Format}
\bibliography{main}


\end{document}